\documentclass[prb,aps,twocolumn,showpacs,floats]{revtex4}
\input psfig.sty

\begin{document}


\title{Transport properties in a simplified double exchange model}

\author{Phan Van-Nham and Tran Minh-Tien}

\affiliation{Institute of Physics,
National Center for Natural Science and Technology, \\
P.O. Box 429, Boho, 10000 Hanoi, Vietnam.
}



\begin{abstract}
Transport properties of Manganite by double exchange mechanism are
considered. The system is modeled by a simplified double exchange model, i.e.
,the Hund interaction between the spins of itinerant electrons and local spins
is simplified to the Ising type. The transport quantities as the electronic
conductivity, thermal conductivity, and the thermal power are calculated by the
dynamical mean-field theory. It is found that the transport quantities
exhibits clearly the ferromagnatic phase transition. A comparison with
experiments is also presented.
\end{abstract}



\maketitle

\section{Introduction}

The interest phenomena in the family of doped manganese oxides
$T_{1-x}D_xMnO_3$ has been recently renewed$^{1-4}$. As doping $x$ and
temperature $T$ are varied, these manganite show a rich variety of
phases$^3$. Particulary interesting is the doping region $0.1<x<0.3$, where
the compounds undergo a transition from either insulating or very high
resistance metalic, paramagnetic (PM) phase at high temperature to a
ferromagnetic (FM) phase at low temperature. Near the transition, the
resistivity of the compounds changes by a orders of magnitude. The application
of a strong magnetic field substantally reduces this effect, thus giving rise
to a very large negative magnetoresistance$^5$. Although the physical
mechanism, responsible for this behaviour has been recently the subject of much
discussion and controversy. The double-exchange (DE) mechanism\cite{Mix}
still  provides a well established sterting point. The DE model was proposed by
Zenner\cite{Mix} who considered the explicit movement of electrons
schematically written as $Mn_{1\uparrow}^{3+}O_{2\uparrow
,3\downarrow}Mn^{4+}\rightarrow Mn^{4+}O_{1\uparrow
,3\downarrow}Mn_{2\uparrow}^{3+}$ where 1,2 and 3 label electrons that belong
either to the oxygen between manganese or to the $e_g$ level of the $Mn$ ions.
In this process, there are two simultaneous motions involving electron moving
from the oxygen to the $Mn^{4+}$ ions and other electron from the $Mn^{3+}$ to
the oxygen$^2$. \\ In the DE process the motion of the itinerant electron
favors the ferromagnetic ordering of the local spins and, vive verse, the
presence of ferromagnetic order facilitates the motion of the itinerant
electron. Hence, only the z-component part of the Hund interaction between the
local spins and spins of the itinerant electron plays an essential role in the
DE. In this paper we study the transport properties by the simplified DE (SDE)
where only z-component part of Hund interaction is incorporated. The transport
quanities such as the dc-conductivity, thermal conductivity and
thermal power are calculated by Dynamical mean-field theory (DMFT). The DMFT
has been extensively used for inverstigating strongly correlated electron
systems. It is based on the fact that the self-energy depends only on frequency
in the infinite dimension limit. Using the DMFT these transport quanities can
be expressed  via the spectral function.\\
We find that the SDE captures main features of the transport properties of
manganites. These results indicate that the DE process in manganites can be
studied by the SDE model which is much simpler than the full version of the DE
model. This provides a starting point tomard to complex including various
variations to the DE mechanism such as the randomness, charge or orbital
ordering.\\
The poster is organized as follows. In section II we present the SDE model and
formulas for dc-conductivity, thermal conductivity and thermal power.
In section III we provide the application DMFT in SDE model. Next section we
present numerical results for the thermal transport illustrating the different
contributions of Hund coupling and concentration of itinerant electrons.
Conclusions are presented in section IV.\\
\section{ Transport coefficents in SDE model} The
Hamiltonian of the SDE model is described as follows\\ \begin{eqnarray}
H=-\sum_{\langle ij\rangle ,\sigma}t_{ij}c_{i\sigma}^{\dagger}c_{j\sigma} -\mu
\sum_{i,\sigma}c_{i\sigma}^{\dagger}c_{i\sigma}-2J_H\sum_{i}S_{i}^zs_i^z,
\end{eqnarray} where $c_{i\sigma}^{\dagger}$ $(c_{\sigma})$ is the creation
(annihilation) operator for an electron itinerant at site $i$ with spin $\sigma
$. The first term includes the hopping only between the nearest neighbour sites
ion, $t_{ij}$ is the hopping integral and is scaled with the spatial dimension
$d$ and to have a finite result in the limit $d\rightarrow \infty $.\cite{mean}
\begin{center} $t_{ij}=\frac{t^{\star}}{2\sqrt d}$
\end{center}
and we take $t^{\star}=1$ as the unit of energy. In the limit $d\rightarrow
\infty$, the bare density of states of the itinerant electron becomes
$\rho (\epsilon)=\frac{1}{\sqrt \pi}e^{-\epsilon^2}$ for a hypercubic
lattice. $\mu $ is chemical potential. The last term is the Hund coupling
between the spins of itinerant $s_{i}^{z}$ electron and local spin $S_i^z$. In
this model, only $z$ component (Ising type) is concerned.
Transport coefficient are calculated with in a Kubo-Greenword formalism,
in which the dc-conductivity $\sigma $, thermal power $S$ and the thermal
conductivity $\kappa $ can be determined from relavant correlation function of
the current operator\cite{Mahan}. We define three transport coefficients as $L^{11}$,
$L^{12}=L^{21}$ and $L^{22}$, we have \\
\begin{eqnarray}
\sigma =\frac{e^2}{T}L^{11}
\end{eqnarray}
\begin{eqnarray}
S=-\frac{1}{eT}\frac{L^{12}}{L^{11}}
\end{eqnarray}
\begin{eqnarray}
\kappa =L^{22}-\frac{(L^{12})^2}{L^{11}}
\end{eqnarray}
where the transport coefficients are found from the analytic continuation of
the relevant "polarization operator" at zero frequency\cite{Mahan}, that
mean\\ \begin{eqnarray}
L^{ij}=\lim_{\nu \rightarrow 0}T Im \frac{L^{ij}(\nu)}{\nu}
\end{eqnarray}
where $L^{ij}(\nu)$ can be calculated from the correlation functions\\
\begin{eqnarray}
{\overline L}^{11}(i\nu_n)=\int_{0}^{\beta}d\tau e^{i\nu_n\tau}\langle
T_\tau {\bf j}(\tau){\bf j}(0)\rangle
\label{eq1}
\end{eqnarray}
\begin{eqnarray}
{\overline L}^{12}(i\nu_n)=\int_{0}^{\beta}d\tau e^{i\nu_n\tau}\langle
T_\tau {\bf j}(\tau){\bf j}_Q(0)\rangle
\label{eq2}
\end{eqnarray}
\begin{eqnarray}
{\overline L}^{22}(i\nu_n)=\int_{0}^{\beta}d\tau e^{i\nu_n\tau}\langle
T_\tau {\bf j}_Q(\tau){\bf j}_Q(0)\rangle
\label{eq3}
\end{eqnarray}
after replace $i\nu_{n}$ by $\nu +i\delta $ ($\delta \rightarrow 0^{+}$), here
$\bf j$ and $\bf j_{Q}$ are particle-current operator and heat-current
operator, respectively. The particle-current operator is defined by the
commutator of the Hamiltonian with the polarization operator $\sum_{i}{\bf
R}_{i}n_i$\cite{Mahan}, with our model we have.\\
\begin{eqnarray} {\bf j}=\sum_{\bf
q}v_{\bf q}c_{\bf q\sigma }^{\dagger}c_{\bf q\sigma}
\label{jj}
\end{eqnarray}
where the velocity operator is $v_{\bf q}=\bigtriangledown_{\bf q} \epsilon (\bf q)$ and
$c_{\bf q\sigma}^{\dagger}$ is Fourier transform of the
$c_{i\sigma}^{\dagger}$: $c_{\bf q\sigma}^{\dagger}=\frac{1}{N}\sum_{\sigma
}e^{i{\bf qR}_j}c_{i\sigma}^{\dagger}$ and the energy-current is defined
by the commutator of the Hamiltonian with the energy polarization operator
$\sum_i{\bf R}_ih_i$ (where $H=\sum_{i}h_i$), with this defined a heat-current
operator is \\
\begin{eqnarray*}
{\bf j}_Q  = {\bf j}_E-\mu {\bf j}
&= &\sum_{{\bf q}\sigma}{\bf
v_q}[\epsilon({\bf q}-\mu ]c_{{\bf q} \sigma}^{\dagger}c_{{\bf q}\sigma }
\end{eqnarray*}
\begin{eqnarray}
-\frac{1}{2}\sum_{{\bf q,q^{'}}\sigma}
J_H\sigma S({\bf q-q')(v_{q'}+v_q)}c_{{\bf q\sigma}}^{\dagger}c_{{\bf q
\sigma}}
\label{jq}
\end{eqnarray}
where $S({\bf q -q'}) = \frac{1}{N}\sum_{i}S_{i}^{z}e^{-i({\bf q -q'}){\bf R}_i}$
The important relations between the heat-current and the particle-current operator are
described fully bellow. \\
Substubting ${\bf j}$ and ${\bf j}_{Q}$ from (\ref{jj}) and
(\ref{jq}) into (\ref{eq1})(\ref{eq2}) and (\ref{eq3}), the transport
coefficients are easy calculated in the infinite dimensional
hypercubic.
\begin{eqnarray} L^{11}=T\sum_{\sigma}\int d\epsilon \rho (\epsilon
)\int d\omega ( -\frac{\partial f(\omega)}{\partial \omega})
A_{\sigma}^{2}(\epsilon ,\omega) \label{l11}
\end{eqnarray}
\begin{eqnarray}
L^{12}=T\sum_{\sigma}\int d\epsilon \rho (\epsilon )\int d\omega (
-\frac{\partial f(\omega)}{\partial \omega}) A_{\sigma}^{2}(\epsilon
,\omega)\omega
\label{l12}
\end{eqnarray}
and\\
\begin{eqnarray}
L^{22}=T\sum_{\sigma}\int d\epsilon \rho (\epsilon )\int d\omega (
-\frac{\partial f(\omega)}{\partial \omega}) A_{\sigma}^{2}(\epsilon
,\omega)\omega^{2}
\label{l22}
\end{eqnarray}
  Above results (\ref{l11}), (\ref{l12}) and (\ref{l22}) appears $A_{\sigma}
(\epsilon ,\omega)$ as spectrum function of its Green function \\
\begin{eqnarray}
A_{\sigma}(\epsilon ,\omega)=-\frac{1}{\pi}ImG_{\sigma}(\epsilon ,\omega)
\end{eqnarray}
That ideal lead us to calculate Green function for each electron with spin
$\sigma $. In this poster, Green function were found for the SDE by DMF theory
approximation. \\
\section{ Application DMF theory in SDE model}
We solve the SDE model (1) by the DMFT. The DMFT is based on the infinite
dimension limit. In the infinite dimension limit the self-energy is pure local
and does not depend on momentum. The Green function of the itinerant electrons
with spin $\sigma$ satisfies the Dyson equation\\
\begin{eqnarray}
G_{\sigma}({\bf k},i\omega_{n})=\frac{1}{i\omega_{n}-\epsilon ({\bf k})+\mu
-\Sigma_{\sigma}(i\omega_{n})},
\end{eqnarray}
So we have local (single-site) Green function\\
\begin{eqnarray*}
G_{L\sigma}(i\omega_{n})&=&\frac{1}{N}\sum_{\bf k}G_{\sigma}({\bf
k},i\omega_{n})
\end{eqnarray*}
\begin{eqnarray}
=\int d\epsilon \rho (\epsilon)\frac{1}{i\omega_{n}-\epsilon ({\bf k})+\mu
-\Sigma_{\sigma}(i\omega_{n})},
\label{g11}
\end{eqnarray}
where $\omega_{n}=\pi T(2n+1)$ is the Matsubara frequency, $\epsilon ({\bf
k})=-2t\sum_{\alpha}\cos(k_{\alpha})$ is the dispersion of the free itinerant
electrons on a hypercubic lattice, $\Sigma_{\sigma}(i\omega_{n})$ is the
self-energy. The self-energy is determined by solving an effective single-site
problem. The effective action for this problem is.\\
\begin{eqnarray*}
S_{eff}&=&-\int d\tau \int
d\tau^{'}\sum_{\sigma}c_{\sigma}^{\dagger}(\tau){\cal
G}_{\sigma}^{-1}(\tau -\tau^{'})c_{\sigma}(\tau^{'})
\end{eqnarray*}
\begin{eqnarray}
+2J_{H}\int d\tau S^{z}\sum_{\sigma}\sigma
c_{\sigma}^{\dagger}(\tau)c_{\sigma}(\tau)
\label{Seff}
\end{eqnarray}
where ${\cal G}_{\sigma}(\tau -\tau^{'})$ is the Green function of the
effective medium ${\cal G}_{\sigma}(i\omega_{n})$ in the time representation
.\\
The local Green function of the effective single-site problem is solely
determined by the partition function. It can be calculted by the equation.\\
\begin{eqnarray}
G_{\sigma}(i\omega_{n})=\frac{\partial {\cal Z}_{eff}}{\partial {\cal
G}_{\sigma}^{-1}(i\omega_{n})}
\label{g22}
\end{eqnarray}
where $Z_{eff}$ is the partition function\\
In addition to (\ref{g11}), the local Green function $G_{L\sigma}(i\omega_{n})$
can be considered as the Green function of a single-site problem with a certain
effective bare Green function ${\cal G}_{\sigma}(i\omega_{n})$ and with the
same self-energy $\Sigma_{\sigma}(i\omega_{n})$, so that we can write\\
\begin{eqnarray}
G_{L\sigma}(i\omega_{n})=\frac{1}{{\cal
G}_{\sigma}^{-1}(i\omega_{n})-\Sigma_{\sigma}(i\omega_{n})}
\label{g33}
\end{eqnarray}
From (\ref{g11})(\ref{g22}) and (\ref{g33}) we have self-consistently
equations, which the self-energy and the Green function are determined. Within
the effective single-site problem, the partition function becomes\\
\begin{eqnarray}
{\cal Z}_{eff}=Tr\int Dc_{\sigma}^{\dagger}Dc_{\sigma}e^{-S_{eff}}
\end{eqnarray}
where the trace is taken over $S_{z}$. This partition function can be
calculated exactly, this similar to DMFT solving the FK model\cite{mean} we
obtain
\begin{eqnarray}
{\cal Z}_{eff}=\sum_{m}e^{-\beta \int d\omega
f(\omega)\frac{1}{\pi}Im\sum_{\sigma ,n}\ln ({\cal
G}_{\sigma}^{-1}(\omega)-J_H\sigma m)}
\end{eqnarray}
where $m =-\frac{3}{2},-\frac{3}{2}+1,\dots \frac{3}{2}$ are projections of
$S$ on $z$ axit.\\
Using Eq.(\ref{g22}) we obtain the local Green function\\
\begin{eqnarray}
G_{L\sigma}(i\omega_{n})=\sum_{m}\frac{w_{m}}{{\cal G}_{\sigma}^{-1}-J_H\sigma
m}
\end{eqnarray}
where
\begin{eqnarray}
w_{m}=\frac{1}{\cal Z}_{eff}e^{-\beta \int d\omega
f(\omega)\frac{1}{\pi}Im\sum_{\sigma ,n}\ln ({\cal
G}_{\sigma}^{-1}(\omega)-J_H\sigma m)}
\end{eqnarray}
\section{Numerical results}
We present the DMFT results for two cases: $n=1.0$ (half
filling) and $ n =0.5$ (quarter filling) where $ n =
-\frac{1}{\pi}T\sum_{\sigma ,n}Im G_{\sigma}(i\omega_{n})$ with two values of
$J=2J_H$ ($J=2$ and $J=4$). The algorithm for determining the Green
function is as follows. ($i$) Begin with the self-energies in each spin
$(\Sigma_{\uparrow }=0.5$ and $\Sigma_{\downarrow}=0.0$). ($ii$) Then
(\ref{g11}) is used to find the local Green function. ($iii$) Subsitituting
$\Sigma_{\sigma}$ in ($i$) and $G_{\sigma}$ were calculated in ($ii$) to
(\ref{g22}) we have the effective medium ${\cal G}_{\sigma}$. ($iv$) Put ${\cal
G}_{\sigma}$ into (\ref{g33}) $G_{\sigma}$ are determined. ($v$) From these
$G_{\sigma}$ and ${\cal G}_{\sigma}$ were determined in ($iii$) and together
with (\ref{g22}) two new $\Sigma_{\sigma}$ are present. Go back to step ($ii$)
and repeat the iteration until convergence is reached. In all our calculations,
the relative error for the Green function of less than $10^{-7}$ is used to stop
iteration loop.
Our results of transport properties in SDE is expressed follow. \\
\begin{figure}[htb]
       \centerline{
\psfig{figure=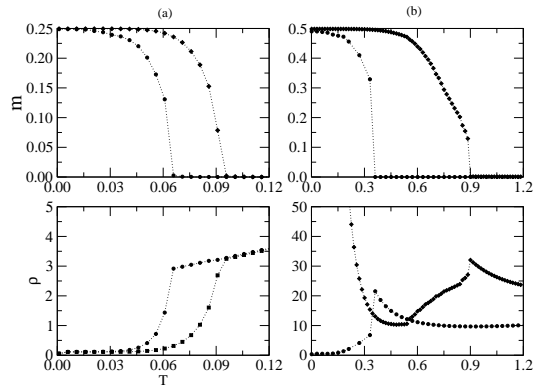,width=0.45\textwidth,angle=-90}
  }
\caption{\small Temperature dependence of the resistivity $\rho$ and magnetization m for
different electron fillings: (a) $(n=0.5)$, (b) $n=1.0$. The line with filled
circles denotes $J=2$ and the line with diamonds denotes $J=4$}
\end{figure}
Fig.1 shows the temperature dependence of the resistivity for different
electron fillings $n$ in two different Hund coupling constant $J$. Fig.1(a)
provides a decrease in the resistivity when the magnetization increases in the
case $n=0.5$. Indeed, in the limit $T\rightarrow 0$, or $m\rightarrow 0.25$,
 the density of states (DOS) of spin-up electron
$A_{\uparrow}(\epsilon ,\omega)=\delta (\omega +\mu -\epsilon)$ is maximum at
$\omega =0$, that mean conduction electron-subsystem becomes a free electron
gas of spin-up electron. Increasing temperature, Fig. 2 shows that, DOS of
spin-up electron is still maximum at $\omega =0$ and system in
ferromagnetic-metalic phase with dependence of resistivity on temperature is
quaratic. On the other hand, in high-temperature paramagnetic phase, $m=0$ as
DOS of spin-up electrons and spin-down electrons is coincides with each other
but their values at $\omega =0$ are not equal zero so system in
paramagnetic-metal phase.\\
\begin{figure}[htb]
       \centerline{
\psfig{figure=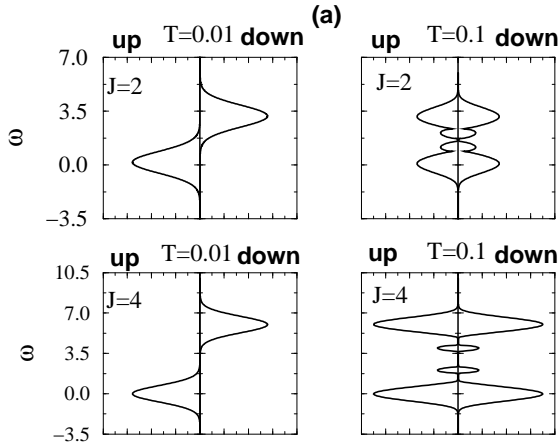,width=0.45\textwidth,angle=0}
  }
\caption{\small Density of states (DOS) $A_{\sigma}(\omega)$ as a function of
frequency for different temperature $T$ in $(n=0.5)$ with different
$J$: left for $J=2$, right for $J=4$. Label 1 and 2 denotes the DOS for
spin-up and spin-down electron, respectively.}
\end{figure}
The sharp decreases of resistivity at $T<T_c$ for
all $J$ is cause by rapid increases of the magnetization $m$. This
discontinuity in the slope $d\rho /dT$ at $T=T_c$ is consequence of the
dynamical mean field approach. Incorporation of spatial spin
fluctuations\cite{Ishi} will smooth out the temperature dependence of the
resistivity in the vicinity of $T_c$, but this is beyond DMFT.\\
\begin{figure}[htb]
       \centerline{
\psfig{figure=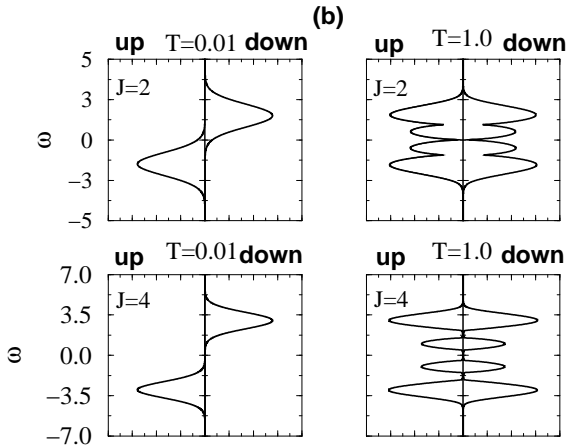,width=0.45\textwidth,angle=0}
}
\caption{\small Density of states (DOS) $A_{\sigma}(\omega)$ as a function of
frequency for different temperature $T$ in $(n=1.0)$ with different
$J$: left for $J=2$, right for $J=4$. Label 1 and 2 denotes the DOS for
spin-up and spin-down electron, respectively.}
\end{figure}
\begin{figure}[htb]
       \centerline{
\psfig{figure=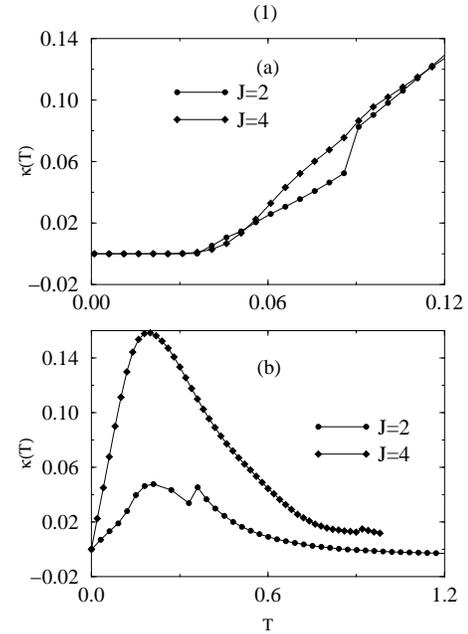,width=0.350\textwidth,angle=0}
}
\caption{\small  Temperature dependence of the thermal conductivity $\kappa $
for different electron fillings: (a) $(n=0.5)$, (b) $n=1.0$. The line with
filled circles denotes $J=2$ and the line with diamonds denotes $J=4$}
\label{h2b}
\end{figure}
\begin{figure}[htb]
       \centerline{
\psfig{figure=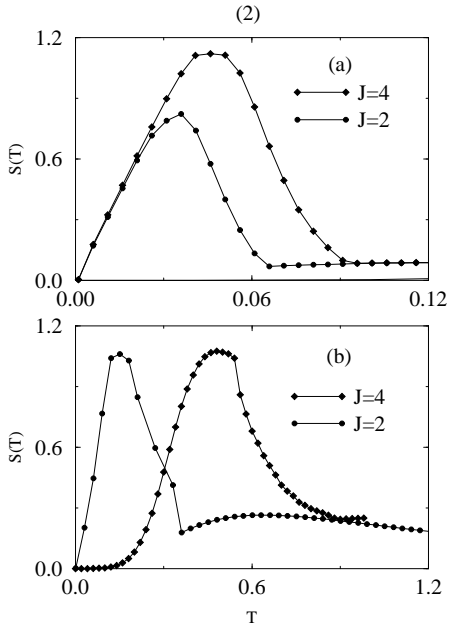,width=0.350\textwidth,angle=0}
}
\caption{\small  Temperature dependence of the thermal power $S$ for
different electron fillings: (a) $(n=0.5)$, (b) $n=1.0$. The line with filled
circles denotes $J=2$ and the line with diamonds denotes $J=4$}
\label{h3a}
\end{figure}
Increasing $J$ constant, the shape of resistivity vs temperature has a litter
modification but the Curie-temperature increases. This analysis is in
qualitative agreement with the calculations in\cite{mas} with disorder
strengths $\bigtriangleup =0$ and the experimental data on manganites.\\
As concentration of itinerant electron increases to $n=1$. Fig. 1(b) shows more
puzzle dependence of function $\rho (T)$ on Hund coupling constant than $n=0.5$
case. Indeed, when $J$ is small ($J=2$), Fig. 3 shows that, at low temperature,
the DOS of spin-up electrons and spin-down electrons overlap each other, that
mean, there is only one band energy so the dependence of resistivity on
temperature is in the metalic phase. But when temperature increases, in
paramagnetic phase, DOS have a gap at $\omega =0$, that mean system in
insulator phase with the negative $d\rho /dT$ above $T_c$. When $J$ is larger
$J=4$, system displays insulating behaviour everywhere as Fig. 3(b) shows in
all temperature, there are a gap at $\omega =0$, except just below the Curie
point, where the rapid increases in the magnetization can cause the resistance
to drop over a small temperature range before it turn around on increases
again. This occurs in the transition from the paramagnetic insulator to the
ferromagnetic-insulator phase because the charge gap in the
ferromagnetic-insulator is smaller than the charge gap in the
paramagnetic-insulator. Those sharp in Fig. 1(b) will generically be smoothed
out by spatial fluctuations.\\
Now, we examine the thermal properties of ours system inclusing the thermal
power $S(T)$ and the thermal conductivity $\kappa (T)$ which are shown in Fig.
4 and Fig. 5.\\
The thermal conductivity behaves as expected with the behaviors of resistivity.
In $n=0.5$ case, system in metalic phase in all temperature so $\kappa (T)$
increases everywhere in temperature range, and when $n=1$ the thermal
conductivity has a sharp at $T_c$ temperature and decreases in insulator
phase.At low temperature, thermal conductivity has a peak and then reach to
zero when $T\rightarrow 0$. This calculations agree with the calculated
electrical thermal conductivity from electrical resistivity in a
$La_{0.67}(Ca,Pb)_{0.33}MnO_3$ single crystal using the Wiedmann-Franz
law\cite{Salamon}.\\
The thermal power behaves as expected with a linear decreases to zero at low
temperature which agrees with the behaviours of $S(T)$ vs temperature for a
mixed $Pb-Ca$ doped sample\cite{Salamon}. At $T_c$ temperature, the slope of
the thermal power $dS/dT$ has a discontinuity, but S(T) does not change sign in
our model, that because in our model the itinerant-electron subsystem is not a
Fermi liquid and the derivative of the chemical potential $d\mu /dT$ does not
change sign at $T=T_c$ (does not agree with prediction in\cite{Gomez})
\section{Conclusions}
  In this poster, we have considered the transport properties in SDE
model by employing DMF theory. With DMF theory we had three self-consistent
equations which is easy calculated by numerical. Although the manganites are
too complicated a system to be described completely by this simple model,
we still arrive at some useful conclusions: The first in this
model the various phases and picture of transition phase of system exit. The
second, qualitative behaviours of $S(T)$ and $\kappa (T)$ agree with experiment
data, special at low temperature. So with this simplified model qualitative
transport properties of DE are presented. But as we have said, above results
only are qualitative conclusion.
Nevertheless, basic transport properties of Maganite are presented and DMF
theory cooperating with DE succeed in investigating Manganite compounds.\\


\begin{thebibliography}{99}
\bibitem{Salamon}M. B. Salamon and M. Jaime, Rev. Mod. Phys. {\bf 73}, 583
(2001). \bibitem{Hotta}E. Dagotto, T. Hotta, and A. Moreo, Physics Reports
{\bf 344}, 1 (2001).
\bibitem{Izy}Y. A Izyumov and Y. N. Skryabin, " Double
exchange model and the unique properties of the manganites" , Physics- Uspekhi
 {\bf 44}, 109 (2001).
 \bibitem{Mix}J. M. D. Coery, M. Viret, and S.Von
Molnar, " Mixed- valence manganites", Advances in Physics. {\bf 48}, 167
(1995).
\bibitem{Kubo}D. M. Edwards, ACM. Green, and K. Kubo, cond-mat/9901133
\bibitem{mean}A. Georges, G. Kotliar, W. Krauth, and M. J. Rozenberg, "
Dynamical mean field theory of strongly correlated fermion systems and the
limit of infinite dimensions", Rev. Mod. Phys. {\bf 68}, 13 (1996).
\bibitem{Kondo}E. Dagotto, S. Yunoki, A. L. Malvezzi, A. Moreo, J. Hu, S.
Capponi, D. Poiblanc, and Furukawa, cond-mat/9709029
\bibitem{hanoi}N.
Furukawa, Proc. Conference on
physics of Manganites (1998), cond-mat/9812066.
\bibitem{mas}B. M. Letfulov and
J. H. Freericks, Phys. Rev. B {\bf 64}, 174409 (2001).
\bibitem{Mahan}G. D. Mahan, " Many-Particle Physics", 2nd edition, Plemum Press
(1990). \bibitem{ba}J. K. Freericks and V. Zlatic, cond-mat/0108500.
\bibitem{Gomez} D. P. Arovas, G. Gomez-Santos, and F. Guinea, Phys. Rev. B
{\bf 59}, 13 569 (1999).
\bibitem{Ishi} S. Ishizaka and S. Ishizara, Phys. Rev. B {\bf 59}, R 8375
(1999).
\bibitem{b}T. Pruschke, D. L. Cox,
and M. Jarrell,Phys. Rev. B {\bf 47} 3553 (1993).
\end{thebibliography}
\end{document}